\documentclass[preprint2]{aastex} 
\usepackage{amsmath} 

\newcommand{\bea}{\begin{eqnarray}} 
\newcommand{\eea}{\end{eqnarray}} 

\newcommand{\nn}{\nonumber} 
\newcommand{\vek}[1]{\boldsymbol{#1}}

\renewcommand{\hat}{ { }} 

\slugcomment{}

\shorttitle{Testing no-hair theorem  with  OJ287} 
\shortauthors{Valtonen et al.}

\begin{document}

\title{Testing black hole no-hair theorem with OJ287}

\author{M.~J.Valtonen\altaffilmark{1,2,3}, S.~Mikkola\altaffilmark{1}, H.~J.Lehto\altaffilmark{1}, 
A.~Gopakumar \altaffilmark{4}, R.~Hudec \altaffilmark{5,6}\and J.~Polednikova\altaffilmark{5,6}} 

\affil{$^1$Tuorla Observatory, Department of Physics and Astronomy, University of Turku, 
    21500 Piikki\"o, Finland}
\affil{$^2$Helsinki Institute of Physics, FIN-00014 University of Helsinki, Finland} 
\affil{$^3$Finnish Centre for Astronomy with ESO, University of Turku}
\affil{$^4$Tata Institute of Fundamental Research, Mumbai 400005,
India}
\affil{$^5$ Astronomical Institute, Academy of Sciences, Fricova 298, 25165 Ondrejov, Czech Republic}
\affil{$^{6}$ Czech Technical University in Prague, Faculty of Electrical
Engineering, Technick 2, 166 27 Praha 6, Czech Republic}

\begin{abstract} 

 We examine the ability to test the black hole no-hair theorem at the $10 \%$ level in this decade using the binary black hole in OJ287.
In the test we constrain the value of the dimensionless parameter $q$ that relates the scaled 
quadrupole moment and spin of the primary black hole: $q_2 = -q\, {\chi^2}$. At the present we can say that $q=1 \pm 0.3$ (one $\sigma$), in agreement with General Relativity and the no-hair theorems. We demonstrate that this result can be improved if more observational data is found in historical plate archives for the $1959$ and $1971$ outbursts. We also show that the predicted $2015$ and $2019$ outbursts will be crucial in improving the 
accuracy of the test.  Space-based photometry is required in 2019 July due the proximity of OJ287 to the Sun at the time of the outburst. The best situation would be to carry out the photometry far from the Earth, from quite a different vantage point, in order to avoid the influence of the nearby Sun. We have considered in particular the STEREO space mission which would be ideal if it has a continuation in 2019 or LORRI on board the New Horizons mission to Pluto.

\end{abstract} 


\section{Introduction} 

 Astronomical observations and detailed astrophysical considerations strongly support
the existence of black hole candidates having masses in the range from few 
$M_{\odot}$ to few $ 10^{10}\, M_{\odot}$. In order to make sure that they actually are black holes as postulated in General relativity (GR), we should prove that at least in one case the black hole no-hair theorems are satisfied.

According to the black hole no-hair theorems, an electrically neutral rotating black hole in GR
is completely described by its mass $M$ and angular momentum $S$
(Israel 1967, 1968, Carter 1970, Hawking 1971, 1972, see Misner, Thorne and Wheeler 1973 for discussions). 
This implies that 
the multipole moments, required to specify the external metric of a black hole, are fully expressible in terms of
$M$ and $S$. In the case of a Kerr black hole, characterized by the Kerr parameter $\chi$, 
its dimensionless quadrupole parameter $q_2$ is  uniquely defined by
\begin{eqnarray}
q_2 = - {\chi^2},
\end{eqnarray}
where $q_2 = {c^4\, Q_2}/{ G^2\,M^3} $ and $\chi = {c\, S}/{ G \,M^2}$, and 
$Q_2$ is the quadrupole moment of the black hole (Thorne 1980; Thorne, Price and Macdonald 1986) .

Recently, the first attempt to probe the black hole no-hair theorems was made by Valtonen et al. (2010) using the available 
optical observations of the BL Lacertae object OJ287. 
The quasiperiodic optical light curve of this quasar \citep{sil88} displays temporal variations having 12 and 60 year cycles (Figure 1).

\begin{figure}
\includegraphics[width=4.5cm,angle=270]{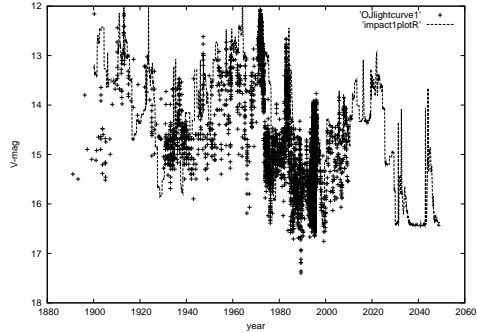} 
 \caption{The optical light curve of OJ287 from 1891 to 2010. The light curve includes previously unpublished data obtained at Harvard by R.Hudec and M.Basta. The line represents the binary black hole model.\label{fig1}}
\end{figure} 

A simple model for OJ287 involves
a secondary black hole orbiting a more massive primary black hole in an eccentric 
orbit having a periodicity of about 12 years, while 
the 60 year period arises from the associated periastron precession \citep{leh96}.
\begin{figure}
\includegraphics[width=4.5cm,angle=270]{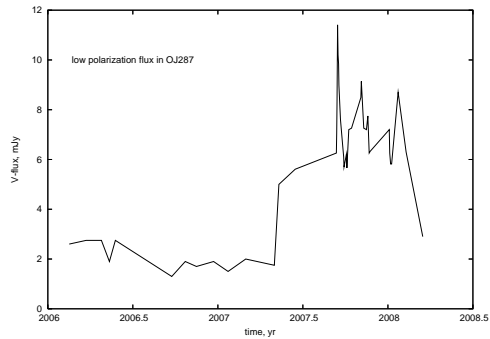} 
 \caption{The optical light curve of OJ287 during 2006-2008. Only low polarization (less than $10\%$) data points are shown. There is a big ``hump'' lasting about one year and a ``spike'' at 2007 September 13 lasting only a few days. It is the ``spikes'' of the light curve that are used to determine the times of impact on the accretion disk and then to calculate the orbit of the secondary.\label{fig2}}
\end{figure}
In principle, one could imagine many different binary models that would satisfy these requirements. However, 
there exists a third requirement which nails down the model. It is related to the observed 
double peak structure in the light curve of OJ287, with the two peaks
separated by one to two years and the pair occuring approximately in every 12 years.
These observations are interpreted as being due to the double impact of 
the secondary black hole on the accretion disk of the primary \citep{leh96}. 
The model was 
able to predict the 2007 September 13 sharp outburst to the accuracy of one day (Valtonen 2007, 2008, Valtonen et al. 2008).
 
 In describing the full light curve of OJ287, one must also calculate the indirect effects of the binary action on the accretion flow. After the orbit is fully determined by the sharp impacts and the related short but bright outbursts, it is straightforward to calculate the more gentle rise and fall of the light curve arising from variations in the accretion flow. Thus the rest of the optical light curve was also predicted with fair accuracy (Sundelius et al. 1997, Valtonen et al. 2009). It is a relatively simple matter to separate these two kinds of flux variations by their quite different time scales (Valtonen et al. 2011, see Figure 2).

These investigations give us the confidence to employ the binary black hole model of OJ287 to test GR.

The first orbit model of Lehto \& Valtonen (1996) made use of 5 outbursts, giving 4 independent intervals of time. They allow a unique solution of 4 orbital parameters, the mass of the primary  $1.71 \pm 0.15 \times 10^{10}\, M_{\odot}$, the eccentricity of the orbit $e= 0.678 \pm 0.004$, the precession rate of the major axis of the orbit $33.3 \pm 2^\circ$, and a constant $\phi_0$ specifying the oriention of the orbit at some initial moment of time.

After the predicted 2005 outburst was observed, a new solution was calculated using 6 outbursts, allowing the determination of 5 parameters (Valtonen 2007). The new additional parameter is the thickness of the accretion disk (scale height $\sim150$ AU), while the precession rate was updated to $37.5^\circ - 39.1^\circ$. The timing of the 2007 outburst together with some new historical data allowed a solution using 9 outbursts, and solving for 8 parameters (Valtonen et al. 2010). These 9 outbursts all follow the basic light curve shape of Figure 3, with a rapid rise to the maximum and then a slower decay to pre-outburst level. The time scales of the outbursts follow the dependence on the impact distance established by Lehto and Valtonen (1996). They form a very well defined sequence. There are no cases when an outburst in this sequence was expected but was not observed. All missing members are at times when there were no observations. Neither are there any extra unexplained members of this sequence.

The new parameters are the spin of the primary black hole, with $\chi_1 = 0.28 \pm 0.08$, the mass of the secondary $1.4 \pm 0.1 \times 10^{8}\, M_{\odot}$, and $q$ which is desribed below.

Parallel to the increase in the number of outbursts in the solution, the number of post-Newtonian (PN) terms was increased in calculating the acceleration between the binary components. Valtonen et al. (2010) include the dominant order general relativistic and classical spin-orbit coupling, which is required in order to relate the dimensionless quadrupole parameter $q_2$ of the primary to its Kerr parameter. They write
\begin{eqnarray}
q_2 = -q\, {\chi^2},
\end{eqnarray}
 and let $q$ be among the 8 parameters of the solution. Its value was determined as  $q=1 \pm 0.3$. Valtonen et al. (2010) noted that the timing of the next outburst in 2015 should 
help to improve the accuracy of the $\chi$ estimate to about $\pm 5 \%$. 
This conclusion along with the fact that the mass of the primary is determined with the 
accuracy of $\pm 1 \%$ prompted us to explore the ways  of testing the no-hair theorem 
at the $\pm 10 \%$ level in {\emph {the current decade}} by measuring $q$ more accurately. 
\begin{figure}[ht] 
\epsscale{1.0}
\includegraphics[angle=270, width=3in] {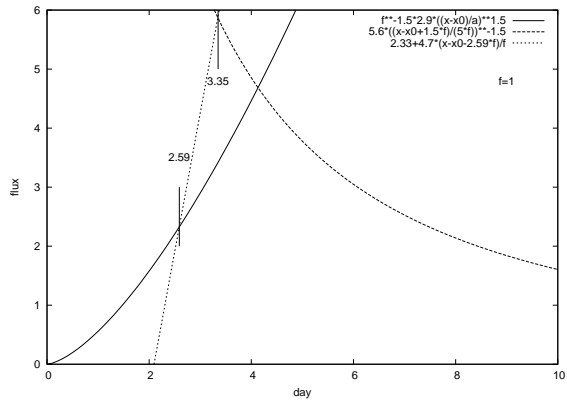} 
\caption{The standard light curve for an OJ287 optical outburst. It is modeled after the unpolarized component of the 2007 September 13 outburst. The model uses three equations in three sections: the formulae are written inside the figure. The parameter $f$ allows applications to slower or faster outbursts since the outburst speed is a function of impact distance measured from the primary black hole. The parameter $a$ has the value 3.
\label{fig3}}
\end{figure} 
 In the literature, there exits a number of proposals to test the black hole no-hair theorems,
plausible in the next decade with the help of electromagnetic and gravitational wave 
observations. The scenarios include radio timing of eccentric  millisecond binary pulsars  
having an extreme Kerr black hole as a companion (Wex and Kopeikin 1999) and 
observing several stars orbiting the massive galactic center black hole at milliarsec distances 
with infrared telescopes capable of doing astrometry at $\sim 10 \mu$arcseconds level \citep{Will2008}. 
Further, LISA observations of gravitational waves from extreme mass ratio inspirals \citep{gb_06} and quasi-normal 
ringdown phases associated with massive black hole mergers \citep{eb_06} will also try to validate 
black hole no-hair theorems.
It has also been argued that the imaging of accretion flow around Sgr A*, if its Kerr parameter is 
not close to one, may allow the testing of the
no-hair theorems in the near future \citep{JP10}. The test relies on the argument that a bright emission ring 
characterizing the flow image will be elliptical and asymmetric if the theorems are violated.  Johannsen and Psaltis (2011) further explore the possibility of detecting modes of quasiperiodic variability in accretion disks as test cases for the no-hair theorems.
  
 In what follows, we briefly summarize our approach, detailed in Valtonen et al. (2010) and list
the improvements desirable for the test. We then identify by timing experiments those impact outbursts (historical and future)
that will be crucial to constrain the $q$ value. Finally, we  
discuss observational requirements to achieve it.

\section{ Additional theoretical inputs and new timing experiments}

\begin{figure}[ht] 
\epsscale{1.0}
\includegraphics[angle=270, width=3in] {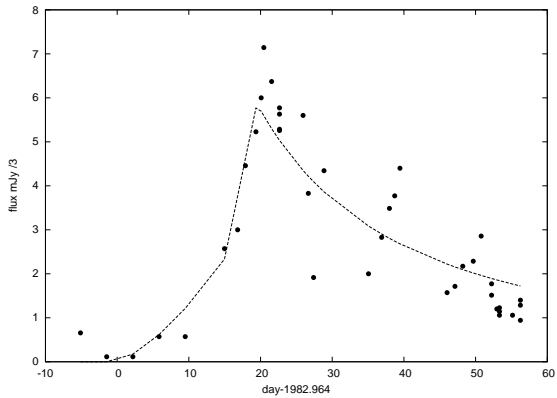} 
\caption{The observations of OJ287 during the 1983 outburst (points). The time scale is in days, with the zero point corresponding to 1982.964. The dashed line shows the standard light curve for an OJ287 optical outburst using $f=6$.
\label{fig4}}
\end{figure} 

In order to obtain the best timing, we construct a 'standard' outburst light curve (Figure 3). It is based on the observations of the 2007 outburst which was monitored throughout the outburst both in total optical brightness and polarization (Valtonen et al. 2008). No such light curve was previously available. These data allow us to separate the underlying Bremsstrahlung emission (unpolarized) from the polarized flares. We model the underlying light curve in three separate sections using analytical functions. From the time zero to time 2.59 days we use a rising power-law form,
\begin{eqnarray}
flux = 2.9 (t/3)^{1.5}f^{-1.5},
\end{eqnarray}
 from day 2.59 to day 3.35 we use a linear rise,
\begin{eqnarray}
flux = 2.33+4.7 (t-2.59f)f^{-1},
\end{eqnarray}
 and beyond day 3.35 we assume a decaying power-law form
\begin{eqnarray}
flux = 5.6((t+1.5f)/(5f))^{-1.5}.
\end{eqnarray}
 The flux is in mJy, and $t$ is the time measured from the beginning of the outburst.

Even though the standard light curve is adopted from observations, the three sections may be justified as follows: At time zero the optical depth of the radiating bubble equals unity and we start see to the interior of the bubble. The optical depth decreases and larger and larger volumes of the bubble come to view. At day 2.59 approximately half of the volume is visible, and thereafter also the rest of the volume produces emission as fast as the visibility front advances into the bubble. This stage happens quickly, in the light travel time of the bubble. At the third stage the flux from the bubble decreases as the radiating plasma cools adiabatically. There is a free parameter $f$ in the formulae which contracts or stretches the outburst time scale. This is necessary since the time scale is a function of the impact distance, measured from the primary black hole. This function is given in Eq. 12 and in Table 3 of Lehto $\&$ Valtonen (1996).

To illustrate the procedure, let us take the light curve of the 1983 outburst. Figure 4 shows the observed points overlaid by the standard light curve. The value $f$ is six, the best fitting value which is also in agreement with Lehto $\&$ Valtonen (1996) time scale. The goodness of fit is judged by minimizing the $\chi^2$. The standard light curve is shifted left-to-right in order to indentify the beginning of the outburst, by minimizing the $\chi^2$ of the difference between observations and the standard curve. This produces a single value for $t_0$, the beginning of the outburst. In order to see how uncertain this value is, we have varied the observed flux values by $\pm1$ mJy in random uniform way. We have also varied the number of points included in the fit, starting from 40 points and adding up to 48 more points to the tail. In this way we get 48 values of $t_0$. Their distribution is centered on 1982.964 and it fits reasonably well with a Gaussian of $\sigma=0.0004$ yr.

\begin{figure}[ht] 
\epsscale{1.0}
\includegraphics[angle=270, width=3in] {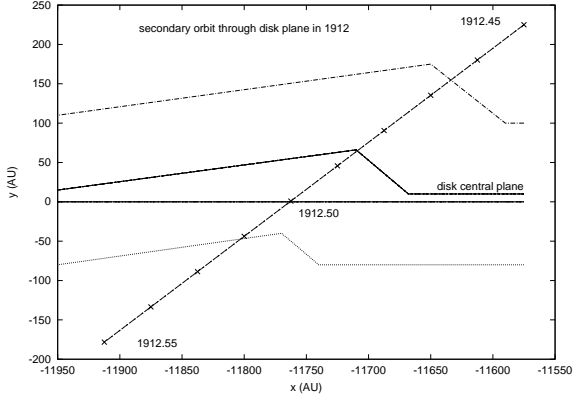} 
\caption{The disk profile during the 1912 impact of the secondary on the accretion disk. The disk central plane is shown, together with two other contours approximately outlining the extent of the disk. The data come from the particle disk simulations of Valtonen (2007). The secondary enters from the upper right and travels to lower left. Its progress is marked by ticks at the interval of 0.01 yr. The x-axis is the radial coordinate measured from the primary, in astronomical units, while the y-coordinate is perpendicular to the disk plane. The original disk was at the zero level in the y-coordinate.
\label{fig5}}
\end{figure} 

The same procedure was applied to other eight outburst light curves. The results are shown in Table 1. Notice that the error limits are generally a little narrower than in Valtonen et al. (2010). In that paper only the first section of the standard light curve was used. Thus the use of the full light curve produces some improvement in timing.

\begin{table}[t] 
\caption{Outburst times with estimated uncertainties. These are starting times of the outbursts\label{outburst}. 
} 
\begin{tabular}{lr} 
1912.970& $\pm$  0.010\\ 
1947.282& $\pm$  0.0005\\ 
1957.080& $\pm$  0.020\\ 
1972.94& $\pm$  0.005\\
1982.964&$\pm$   0.0005\\  
1984.130&$\pm$   0.002\\ 
1995.843&$\pm$   0.0005\\ 
2005.74& $\pm$  0.005\\ 
2007.692&$\pm$   0.0005\\ 
\end{tabular} 
\end{table}
  
We may also ask how accurately the theoretical model can determine the outburst timing. In the impact model a typical disk crossing time of the secondary black hole is one week. However, events at much shorter time scale can make a difference to the outburst timing, as illustrated by Ivanov et al. (1998). They show that there is a factor of two pressure change over the distance of 1/30 of the disk width ahead of the secondary black hole. Thus we may consider time steps of $\sim 1/30$ of a week as physically meaningful, i.e. the relevant time step is $\sim 0.0005$ yr. Also in the gas bubble bursting out of the disk this same time scale produces a significant amount of evolution, as shown in their Figure 4. We consider this the minimum time step that has astrophysical relevance. 

Another limitation to the timing accuracy comes from the influence of the secondary on the level of the accretion disk. The approaching secondary lifts the disk up, and cause an impact earlier than predicted in a rigid disk model (Ivanov et al. 1998). The calculation of this effect by particle disk simulations was carried out by Valtonen (2007). In Figure 5 we show the profile of the accretion disk immediately after the impact in the summer of 1912. We notice that the disk is lifted towards the approaching secondary and is bent. In the model we need to know the raised level of the accretion disk at the time of the impact. The orbit of the secondary is marked by ticks at the intervals of 0.01 yr. We estimate that the timing accuracy in this occasion is $\pm 0.005$ yr. This is the typical accuracy that we may use for impacts at the outer disk ($i.e.$ the 1913, 1957, 1973, 2005, 2015 and 2022 outbursts). For the inner disk the effect is negligible. For example, the 2005 outburst may be timed within $\pm0.001$ yr from observations, but such accuracy is not justified by theoretical considerations.

The PN approximation provides the equations of motion
of a compact binary as corrections to the Newtonian equations of motion
in powers of $(v/c)^2 \sim G M / (c^2 R)$,
where $v$, $M$, and $R$ are
the characteristic orbital velocity,
the total mass, and the typical orbital separation of the binary,
respectively. In Valtonen et al. (2010), the binary black hole was modeled using a spinning primary
black hole with an accretion disk and a non-spinning companion. The calculation of the orbit included all the 2PN-accurate non-spinning finite mass contributions as well as 
the leading order general relativistic (1.5PN order) and classical spin-orbit (2PN order) spinning contributions, and radiation reaction effects (2.5PN order).
Here 
the terminology 2PN, for example, refers to corrections to Newtonian dynamics in powers of $(v/c)^4 $.

For the present work, we incorporated the following three new features in orbit calculation.

    The non-spinning finite mass contributions to the binary black hole dynamics are now fully 3PN accurate. 
This is achieved by adding the 3PN contributions to $d^2{ {\vek x}}/dt^2 $, where 
$\vek x$ is the relative separation vector in the center-of-mass frame, 
as given by Eq.~(1) in  Valtonen et al. (2010b). These 3PN contributions are available in Mora \& Will (2004).
Secondly, we let the smaller black hole also spin. Therefore we add the leading 
order spin-spin contributions to $\ddot { {\vek x}}$, given by Eqs.~(54) in  Barker \& O'Connell (1979). These contributions appear at the 2PN order. This is a desirable addition due 
to the fact that the classical spin-orbit coupling,
crucial to constraining the value of $q$, also enters orbital dynamics at the 2PN order.
Furthermore, we have also included the contributions due to classical spin-orbit and 
general relativistic spin-spin interactions to the precessional equation for 
the unit spin vector $\vek s_1$
in the present analysis.
The compact binary dynamics, employed in the present investigation, schematically reads
\begin{eqnarray} 
\ddot { {\vek x}} \equiv 
\frac{d^2  {\vek x}} { dt^2} &=& 
\ddot { {\vek x}}_{0} + \ddot { {\vek x}}_{1PN} \nonumber 
+ \ddot { {\vek x}}_{SO}+  \ddot { {\vek x}}_{SS} +  \ddot { {\vek x }}_{Q}\\ 
&& + \ddot { {\vek x}}_{2PN} +  \ddot { {\vek x}}_{2.5PN} + \ddot { {\vek x}}_{3PN} \,,  \\
\frac{d {\vek s}_1}{dt} &=& \left (  {\vek \Omega}_{SO} + {\vek \Omega}_{SS} + {\vek \Omega}_{Q} \right ) \times {\vek s}_1 \,,  
\end{eqnarray} 
The explicit expressions for the non-spinning contributions to $\ddot {\vek x} $ are listed in  Mora \& Will (2004).
The spin related contributions to $\ddot { {\vek x}} $ and $ d {\vek s_1}/dt $  are from Barker \& O'Connell (1979).
The additional spin related contributions to the dynamics, namely 
$\ddot { {\vek x}}_{SS}$, $ {\vek \Omega}_{SS} $ and $ {\vek \Omega}_{Q} $, that are 
not listed in  Valtonen et al. (2010a), are given by
\begin{eqnarray}
\ddot { {\vek x}}_{SS} &=& - \left (\frac{3\, G^3\,m^3}{c^4\, r^4} \right )\,
\chi_1\, \chi_2\, \eta 
\biggl \{  
\left ( {\vek s}_1 \cdot \hat {\vek n} \right ) \, \vek s_2
\nn \\ &&
 +
\left ( {\vek s}_2 \cdot  \hat {\vek n} \right ) \, \vek s_1  -
5\, \left ( {\vek s}_1 \cdot  \hat {\vek n} \right )  
\left ( {\vek s}_2 \cdot  \hat {\vek n} \right ) \hat {\vek n}
\nn \\ &&
+ \left ( {\vek s}_1 \cdot  {\vek s}_2 \right ) \, \hat {\vek n}
\biggr \}\,,\\
{\vek \Omega}_{SS} &=&
\left ( \frac{G^2\,m_2^2}{c^3\, r^3} \right )\,\chi_2 
\biggl \{ 
3\, \left (  {\vek s}_2 \cdot  \hat {\vek n} \right ) \hat {\vek n}   -  {\vek s}_2 
\biggr \}\,,\\
{\vek \Omega}_{Q} &=&
\left ( \frac{G^2\,m^2\, \eta}{c^3\, r^3} \right ) \, q\, \chi_1 
\biggl \{ 
3\, \left (  {\vek s}_1 \cdot  \hat {\vek n} \right ) \hat {\vek n }  -  {\vek s}_1 
\biggr \}\,,
\end{eqnarray}
where the Kerr parameter $\chi_1$ and the unit vector
${\vek s}_1$ define the spin of the primary black hole by the relation
${\vek S}_1 = G\, m_1^2 \, \chi_1 \, {\vek s}_1/c$, while
 $\chi_1$ is allowed to take values between $0$ and $1$ in GR.
A similar rule applies to $\chi_2$ in ${\vek S}_2 = G\, m_2^2 \, \chi_2 \, {\vek s}_2/c$.
The vector $\hat {\vek n}$ is defined to be
$ \hat {\vek n} \equiv {\vek x}/r $, where 
$ r = | {\vek x} |$, while $m= m_1 +m_2 $ and 
$\eta = m_1\, m_2/m^2$.

The effect of including the 3PN corrections to the orbital dynamics is roughly 
a one percent increase in the estimated mass of the bigger black hole, demonstrating that the 
employed PN dynamics is in the convergent regime (Valtonen et al. 2010b).
Therefore, its influence on the $q$ estimate is within our desirable error limits.
\begin{figure}[ht] 
\epsscale{1.0}
\includegraphics[angle=270, width=3in] {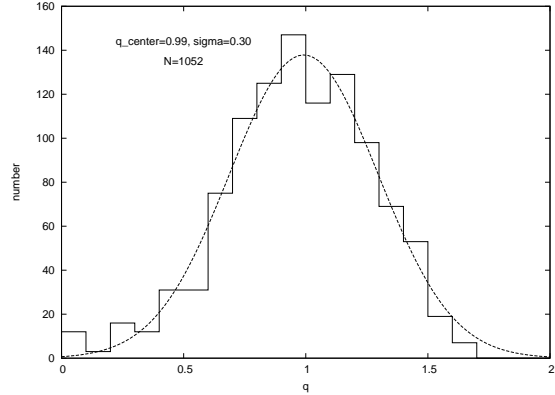} 
\caption{The distribution of the final $q$-values among the solutions using the currently best impact timings. We combine sets 7 $\&$ 8 to create the $q$-distribution. 
\label{fig6}}
\end{figure} 

The fact that the primary black hole is spinning slowly,
in other words $\chi_1$ is much less than 1, indicates
that the spin-spin contributions to
$\ddot { {\vek x}}$ enter the binary dynamics at the 3PN order.
Note that it is the definition of the spin of a compact object, namely
${\cal S} \sim G \, m_{\rm
co}^2 \, v^{\rm spin} /c^2 $, where $m_{\rm co}$ and
 $v^{\rm spin}$ are the typical mass and
rotational velocity of the spinning compact object,
that makes the spin-spin contributions to appear
at the 3PN order in our model.
Further,
the presence of the symmetric mass ratio $\eta$
as a common factor in the spin-spin corrections (
 $\eta \sim  10^{-3}$ in our binary black hole
model) ensures that these corrections have
 only minor effects on the
orbits.

The combined effect of higher PN order and the presence of 
$\eta$ as a common factor makes sure that the leading order spin-spin and 
classical spin-orbit couplings make negligible contributions to 
$\dot {\vek s}_1$.  The timing experiments also reveal that the change in the 
orientation of the secondary spin axis does not affect the $q$ estimates.

 In this investigation, we make a third improvement based on astrophysical considerations. 
In Valtonen et al. (2010a), the black hole spin of the primary black hole 
was parallel to the accretion disk spin
at the initial epoch, which was the year 1856. Due to PN effects, 
the black hole spin wanders about $9^{ \circ }$  off from this direction during its precession cycle
that lasts around 1300 years. 
In the present model,  the precession cone axis coincides with the mean accretion disk axis. 
After performing a number of numerical experiments, we found that it is possible to choose 
a suitable initial direction for 
 $\vek s_1$ such that 
the angle between the spin and the disk axes remains constant ( $ \sim 8^{\circ}$) during the precessional motion of $\vek s_1$.

 It is 
reasonable to expect such a situation due to the Bardeen-Peterson effect. Because the time scale of the Bardeen-Peterson effect is much longer than the black hole spin 
precession time scale (Lodato and Pringle 2006), the two directions do not coincide.
 The time scale of the Bardeen-Peterson effect is of the order of one million years  (Natarajan and Pringle 1998, Eq. 2.16) which is intermediate between the spin precession time scale of $10^3$yr and the binary merger evolution time scale of about $10^8$ yr (Iwasawa et al. 2011). Thus we expect that in $10^8$ yr the Bardeen-Peterson effect is important up to the distance of about $10^2$ Schwarzschild radii in the disk (Natarajan and Pringle 1998, Eq, 2.8), but the disk can follow only the mean direction of the spin. It cannot keep up with the $10^3$ yr evolution of the actual spin.

\begin{figure}[ht] 
\epsscale{1.0}
\includegraphics[angle=270, width=3in] {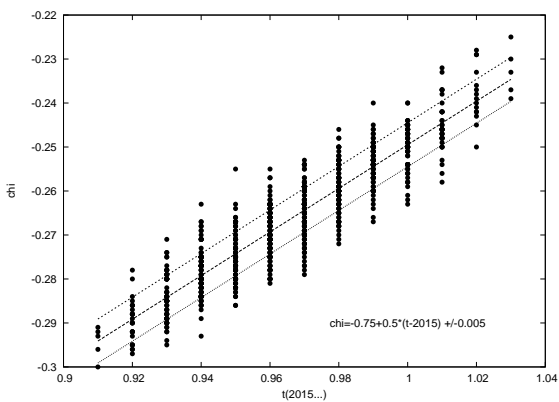} 
\caption{The correlation of $\chi_1$ with the zero point $t_0$ of the 2015 outburst (labelled $t(2015...)$. The insert gives the best fit functional dependence. Set 3.
\label{fig7}}
\end{figure} 

 With the above mentioned additional features, we have searched for orbit solutions. As before, an automatic search algorithm is used. It takes a trial orbit, then improves it until all nine outbursts happen within their allotted time intervals. Typically one solution is found in 3 minutes of computing time with a modern PC. We have used sets of 1080 orbits with given standard parameters. However, the convergence was not always found in a reasonable amount of time. Then the attempt to find a solution was discarded and the next trial was started. For this reason the number of orbits in a set is always less than 1080.

In Table 2 we give the set number, the orbit number and value of the dimensionless spin of the primary $\chi_1$ in the first three columns, respectively. The spin value was generally taken as  $\chi_1=0.275$, except in two sets (3 \& 11) where a range of $\chi_1$ values were used. The next column in Table 2 gives the value of the secondary spin. The spin $\chi_2$ components are either -0.5,-0.5,-0.5 (standard case), 0,0,0 (set 6) or +0.5,+0.5,+0.5 (set 5). Smaller sets were calculated to ascertain that these three $\chi_2$ values are representative in statistical sense of the different orientations and magnitudes of $\vek s_2$. The last column in Table 2 gives the range of the parameter $q_0$
which is initially uniformly distributed between the limits. The solutions converge to a distribution of $q$ which is narrower than this range. Only in set 3 a fixed value of $q_0=1$ was used.

Even though $q_0$ is not a physical parameter but an ingredient of the orbit finding algorithm, its proper choice is still important. We tried setting $q_0$ initially far from the value $q_0=1$, using either $q_0=0$ or $q_0=2$, but we found that our code was not able to find enough solutions to justify these choices. For example, in the latter case only 23 solutions were found which concentrate around $q_{center}=1.16$ with a standard deviation of 0.15. Taking the distribution uniformly between these two limits produces more solutions, but mostly from the range between $q_0=0.6$ and $q_0=1.4$. Therefore we decided to carry out most experiments using this range of $q_0$. However, since it is possible to add some solutions also using the wider range of $q_0$, we have sometimes added two sets together, one with the narrower range, and the other one with the wider range. The distribution of $q_0$ then mimics a Gaussian with the standard deviation of 0.42. The resulting $q$ distribution is always narrower than this, demonstrating that we are not biasing the final $q$ distribution to be unduly compact by our choice of $q_0$.   

There were additional conditions in some sets which are not listed in Table 2. In set 4 the outburst uncertainty limits were taken from Valtonen et al. (2010). They are generally somewhat wider, and also some of them are centered a little differently from the ranges listed in Table 1. (Note that Valtonen et al. 2010 has a misprint; one of the central values they use is 1995.843, not 1995.841). On the other hand, sets 7 and 8 explore the solutions where one of the intervals is made narrower, i.e. the range of the timing of the 2005 outburst is $2005.74\pm0.001$, five times norrower than in our standard case. In general the initial angle between the disk and the primary spin $\chi_1$ is $8^\circ$, but in set 10 we tested also the case of an initially zero angle. In sets 11 and 12 the $t_0$ in 1995 is shifted down and up by 3.5 hours, respectively.
    
\section{Results}

For every set we have constructed the distribution of $q$ values, and since these distributions resemble a Gaussian, we have determined the best fitting Gaussian parameters, the central value $q_{center}$ and the standard deviation $\sigma$ for each distribution. These are listed in Table 3, together with the errors in each parameter. Figures 6 illustrates one such distribution, a combination of sets $7\&8$. 
\begin{figure}[ht] 
\epsscale{1.0}
\includegraphics[angle=270, width=3in] {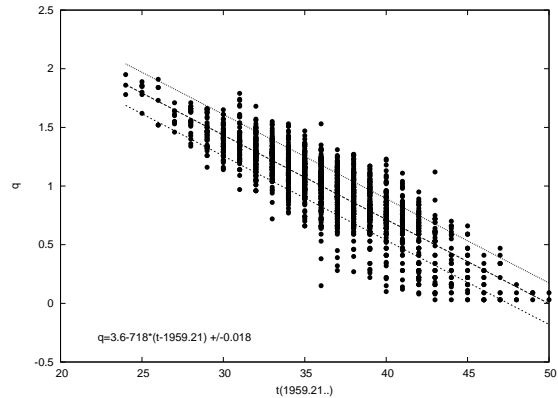} 
\caption{The correlation between $q$ and the starting time of the 1959 outburst $t_0$ (labelled $t(1959.21..)$ when $\chi_1=0.275$. The functional form of the correlation is given inside the figure. There is also a correlation with the $\chi_1$ such that the line of regression is shifted to the right in the figure by one unit for a decrease of $\chi_1$ by 0.015 units. Five units in the time axis corresponds to 4.4 hours.
\label{fig8}}
\end{figure}

\begin{table}[h]
\caption{Parameters of the sets of solutions.
}
\begin{tabular}{lcccl}
Set& No.& $\chi_1$   & $\chi_2$    & $q_0$  \\ 
 1&1012& 0.275 & -0.87 & 0.6-1.4     \\
 2& 864& 0.275 & -0.87 & 0.0-2.0      \\
 3& 901&0.26$\pm$0.04 & -0.87 & 1.0     \\
 4& 362& 0.275 & -0.87 & 0.6-1.4 \\
 5&1009& 0.275 & +0.87 & 0.6-1.4      \\
 6&1017& 0.275 & 0.0 & 0.6-1.4      \\
 7& 598& 0.275  & -0.87 & 0.6-1.4 \\
 8& 454& 0.275 & -0.87 & 0.0-2.0 \\
 9& 914&0.26$\pm$0.03& -0.87 & 0.6-1.4       \\
10& 283&0.27& -0.87 & 0.6-1.4       \\
11& 658&0.275& -0.87 & 0.6-1.4       \\
12&1015&0.275& -0.87 & 0.6-1.4       \\

\end{tabular}
\end{table}

The Kerr parameter of
the primary black hole $\chi_1$ will be constrained by the timing of the outburst in 2015 (Valtonen et al. 2010a). Figure 7 shows the correlation of $\chi_1$ with $t_0$, the zero point of the 2015 outburst, using set 3. The accuracy of the $\chi_1$ determination, after the the 2015 outburst time is known, is $\pm0.005$ (one $\sigma$). It is likely that OJ287 becomes a "Christmas star" of 2015.

We will now discuss those historical outbursts which were not employed in finding the orbital 
solution. We will look for a correlation between the starting times of these outbursts with $q$. We will also ask whether the distribution of $q$ can be narrowed down by future observations. 
    
\begin{table}[h]
\caption{Gaussian fits to $q$ distributions.
}
\begin{tabular}{lcccl}
Set & $q_{center}$&error   & $\sigma$&error    \\ 
1 & 1.00 & 0.01&0.26&0.01\\
2 & 1.01 & 0.05&0.59&0.04\\
3 & 1.03 & 0.01&0.08&0.01\\
4 & 1.01 & 0.02&0.31&0.02\\
5 & 0.96 & 0.01&0.24&0.01\\
6 & 0.98 & 0.01&0.25&0.01 \\
7 & 0.98  & 0.01&0.25&0.01 \\
8 & 0.98 & 0.02&0.42&0.02\\
9 & 1.00 & 0.01&0.27&0.01\\
10 & 0.58 & 0.01&0.19&0.01\\
11 & 0.74 & 0.01&0.19&0.01\\
12 & 1.06 & 0.01&0.30&0.01\\
1+2 & 1.02 & 0.01&0.33&0.01\\
1+2+5-9 & 0.99 & 0.01&0.12&0.01\\
\end{tabular}
\end{table}

\begin{figure}[ht] 
\epsscale{1.0} 
\includegraphics[angle=270, width=3in] {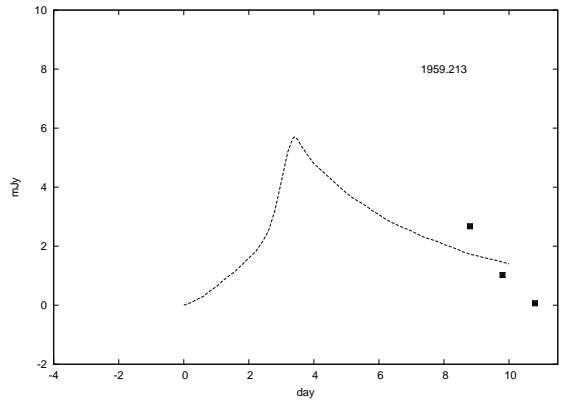} 
\caption{The observation of the brightness of OJ287 at the expected 1959 outburst time. The dashed line is the template from the 2007 
outburst while the squares reprent the three observations. The base level has been arbitrarily normalised 
as there are not enough data to determine it.\label{fig9}} 
\end{figure} 

The first outburst which we studied is the 1934 outburst, with the expected starting time 1934.3439 if $\chi_1=0.275$ and $q=1$. The correlation of the starting 
time with $q$ is so weak that this outburst is of no interest in determining $q$. Moreover, there are no data in the historical light curve yet to verify this outburst.
The 1935 outburst is no better in this respect. It is expected at 1935.3939, with little correlation with $q$. This outburst has not been verified either in the observational record.
\begin{figure}[ht] 
\epsscale{1.0}
\includegraphics[angle=270, width=3in] {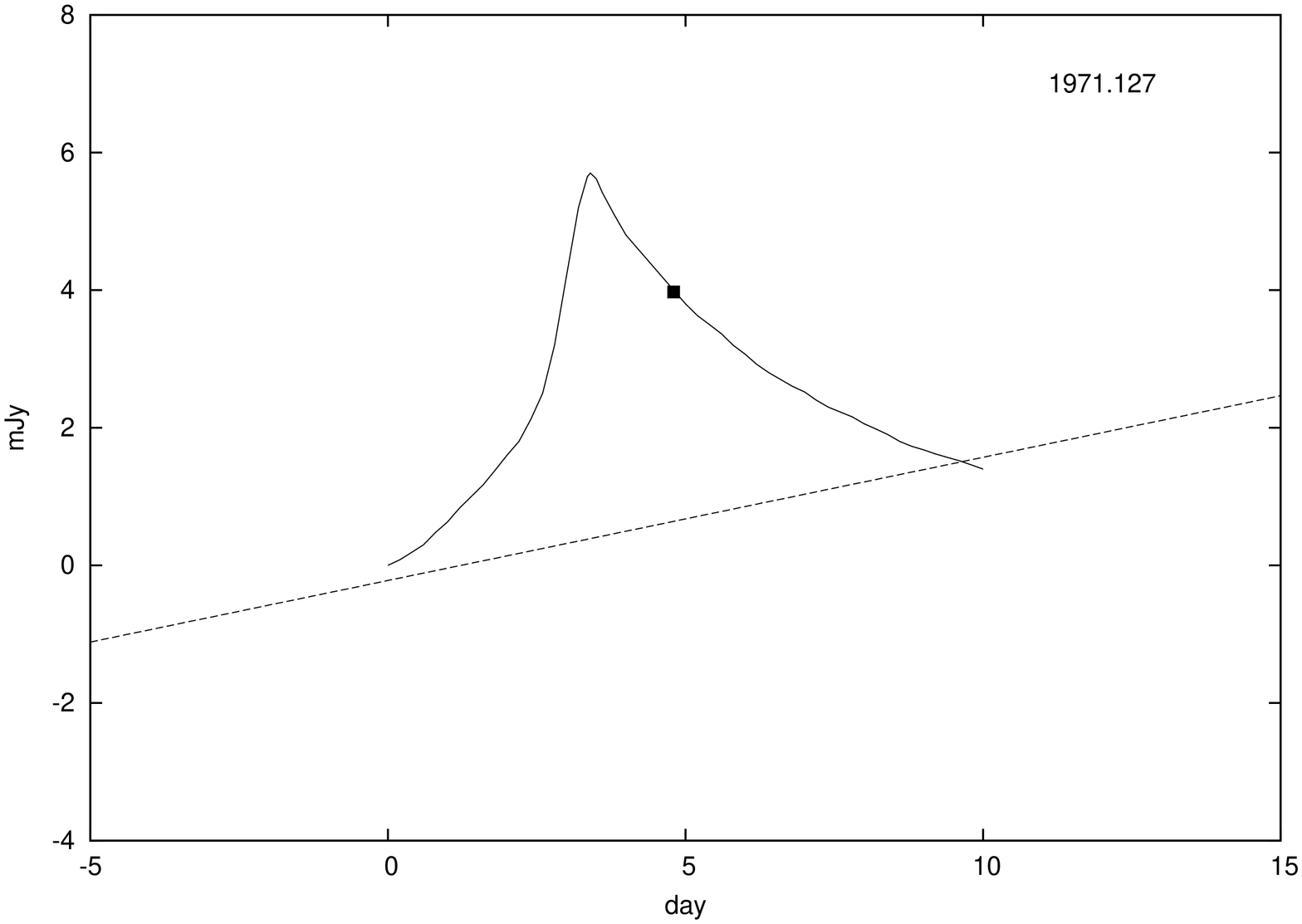} 
\caption{Observations of the brightness of OJ287 at the expected 1971 outburst time. The solid line is the template from the 2007 
outburst while the square represents the sole observational point. The base level (dashed line) is uncertain.
\label{fig10}}
\end{figure} 

The next interesting outburst should have taken place in 1959. Here we expect quite a strong correlation between the start of the outburst and $q$ (Figure 8).

\begin{figure}[ht] 
\epsscale{1.0}
\includegraphics[angle=270, width=3in] {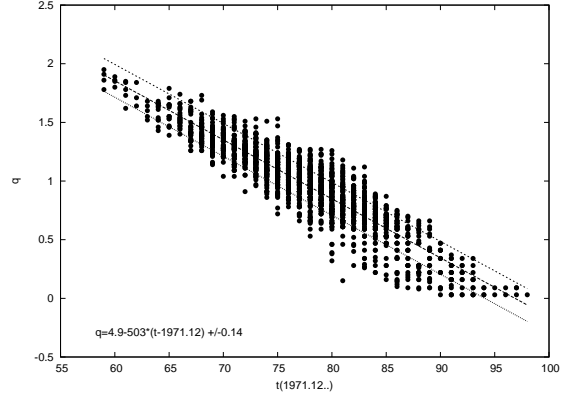} 
\caption{The correlation between $q$ and the starting time of the 1971 outburst $t_0$ (labelled $t(1971.12..)$ when $\chi_1=0.275$. The functional form of the correlation is given inside the figure. There is also a correlation with the $\chi_1$ such that the line of regression is shifted to the right in the figure by one unit for a decrease of $\chi_1$ by 0.01 units. Five units in the time axis corresponds to 4.4 hours.
\label{fig11}}
\end{figure}

We see from Figure 8 that the difference in the starting time of the outburst by 15 units corresponds to the range of 0.8 in $q$. Since 15 units in the figure corresponds to 13 hours, with good timing, say with 5 hour accuracy, it should be possible to determine the $q$ value at the level of $ 1 \pm 0.20$ if a fair number of detections are found in the historical plate 
collections. The current observational situation is 
depicted in Figure 9.

The next outburst of interest is the 1971 outburst and at present there exists 
only one observing point (Figure 10). Figure 11 shows the expected correlation between $q$ and $t_0$. If enough observations are found, it should be possible to determine the $q$ value with the accuracy as high as $ 1 \pm 0.16$ units.

\begin{figure}[ht] 
\epsscale{1.0} 
\includegraphics[angle=270, width=3in] {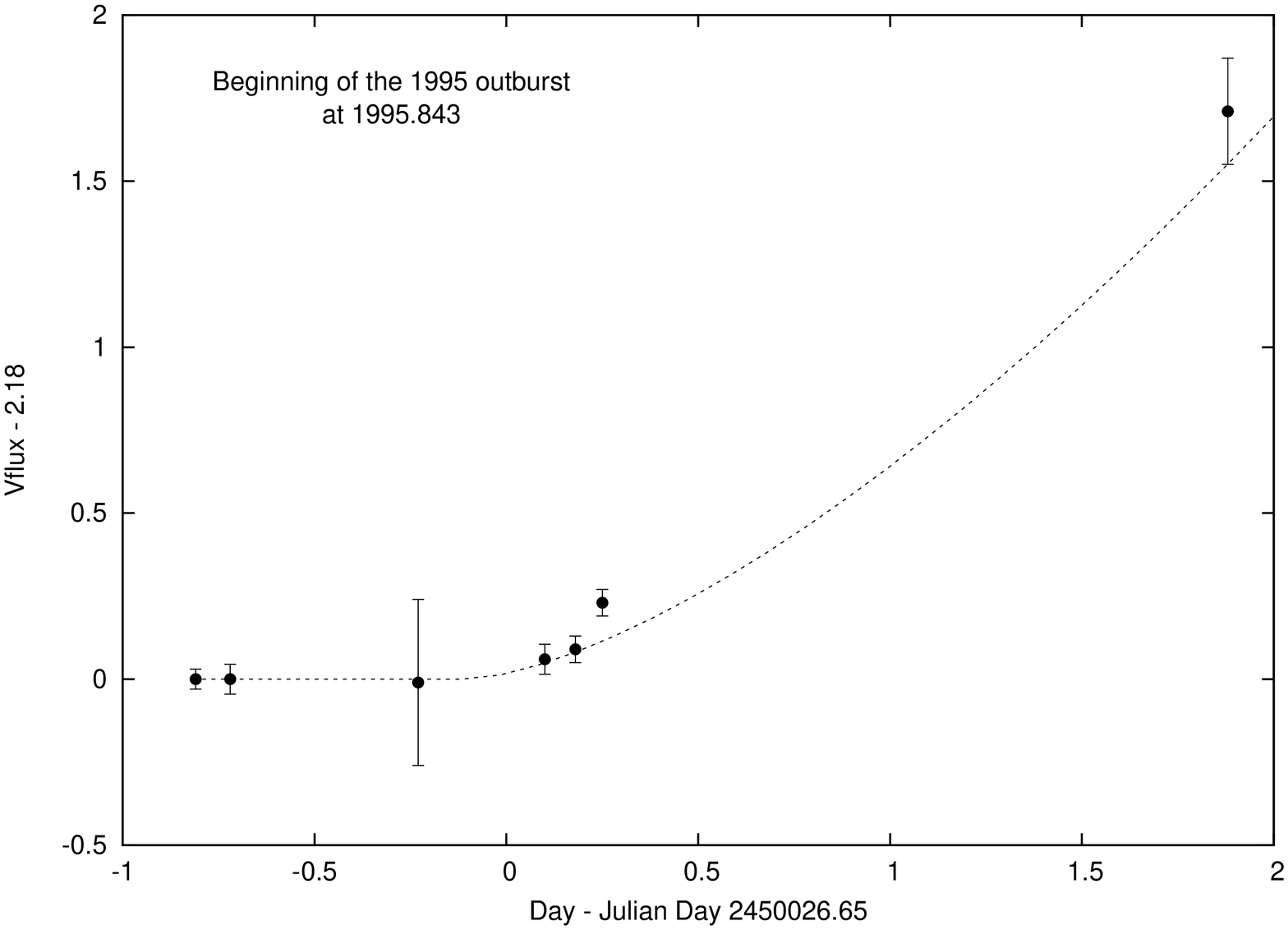}
\caption{ 
 The observations of OJ287 at the beginning of November 1995, transformed to optical V-band. Altogether 50 observations have been binned to 7 points. Overlaid is the theoretical light curve profile from Figure 3. The zero point of time is
at Julian Day (JD) 2450026.65, i.e at 3:36 hours GMT on November 6, 1995.\label{fig12} } 
\end{figure}

The 1995 outburst was already used in our solution. There was an intensive monitoring campaign of OJ287 (called OJ94) during this outburst season, 
but unfortunately there exists a gap in these observations just at the crucial time (Figure 12). 
It may still be possible that there are measurements somewhere which are not recorded in the OJ94 campaign light curve, 
and which would be valuable in narrowing down $q$ even from these data. The line in Figure 12 is drawn using the standard light curve of Figure 3 as a template to compare with the 1995 observations. In set 11 the value of $t_0$ has been shifted down by 3.5 hours, and in set 12 it is shifted up by the same amount. The shifts lead to shifting $q_{center}$ up or down by $\sim 15\%$. It should be noted that 
even a few more measurements of 1995 could narrow down the range of $q$.

 \begin{figure}[ht] 
\epsscale{1.0}
\includegraphics[angle=270, width=3in] {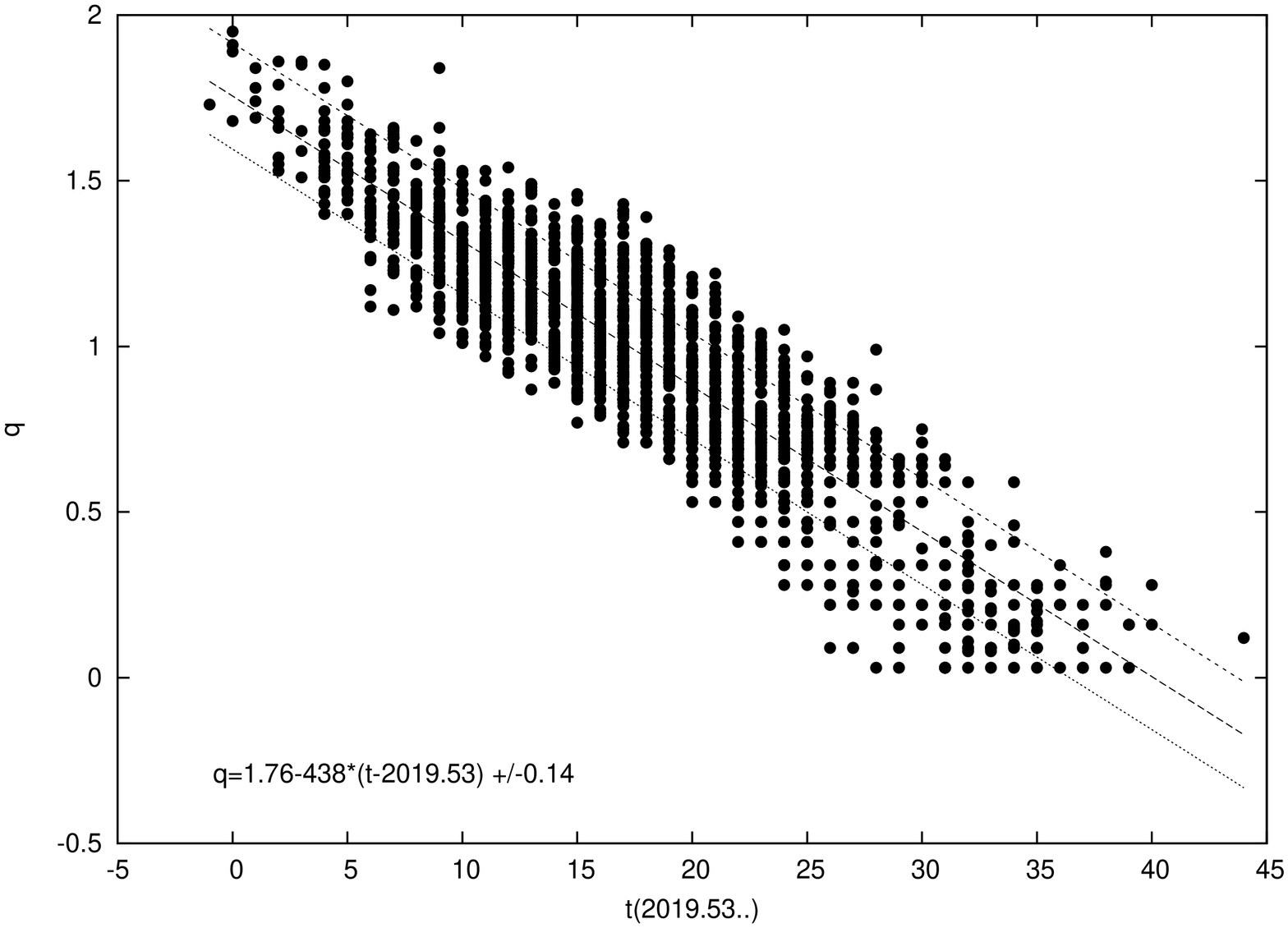} 
\caption{The correlation between $q$ and the starting time of the 2019 outburst $t_0$ (labelled $t(2019.53..)$ when $\chi_1=0.275$. The functional form of the correlation is given inside the figure. There is also a correlation with the $\chi_1$ such that the line of regression is shifted to the right in the figure by one unit for a decrease of $\chi_1$ by 0.0125 units. Five units in the time axis corresponds to 4.4 hours.
\label{fig13}}
\end{figure}


  Let us now turn our attention to the expected future outbursts in our binary black hole model.
We expect three more outbursts during the next two decades occuring in 2015, 2019 and 2022. 
As we mentioned above, the 2015 outburst should be an easy one to detect, as it is expected in December of that year. 
The exact date will 
in fact give us a good spin value. 
The dependence on $q$ is secondary, and thus it is of no use by itself for the testing of the no-hair theorems.

  The 2019 outburst is sensitive to the $q$ and with $\chi = 0.275$, 
it should begin at 2019.53175 if $q=1$ (Figure 13). With good timing the $q$ value is determined with the accuracy of $1 \pm 0.16$ (Figure 14). If by good luck we will find the necessary historical data to time both the 1959 and 1971 outbursts in addition to observing the 2019 outburst, we get close to the $10\%$ accuracy in $q$ (see the combined sets 1+2+5-9 in Table 3). Improvements in understanding the astrophysical processes in OJ287 may then bring the accuracy even below $10\%$. Without the accurate timing in 2019, the $q$ value cannot be determined better than to  $1 \pm 0.3$ even if the $\chi_1$ determination in 2015 is a success (Figure 6).

\begin{figure}[ht] 
\epsscale{1.0}
\includegraphics[angle=270, width=3in] {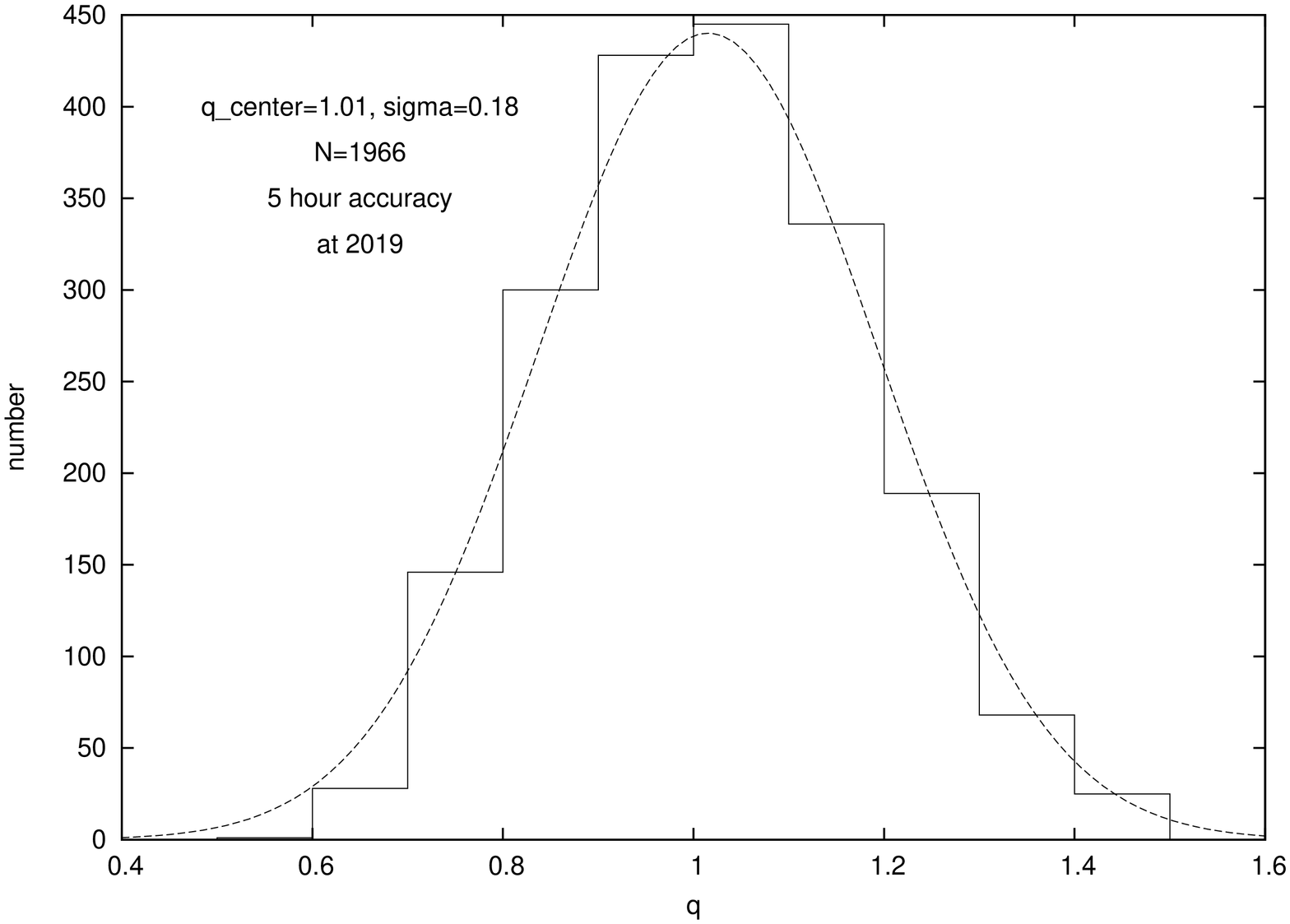} 
\caption{The distribution of $q$ in the simulations where the 2019 outbursts has been timed with the accuracy of 5 hr and also the $\chi_1$ value has been found by the observations of the 2015 outburst. This gives the practical limit of the method of using OJ287 to test General Relativity if historical records do not bring additional information at key moments of time.
\label{fig14}}
\end{figure}

However, observing OJ287 during the expected 2019 outburst window, namely around  2019 July 21, is 
practically impossible from the ground.
This is because the angular distance between the Sun and OJ287 
in the sky is only about 12 degrees at the beginning of this event, and it goes down to 8 degrees by the time of the peak flux. 
It may require space observations to carry out the measurement. 
The 2022 outburst is scheduled practically at the same time of year as the 2019 outburst. 
Obviously it would also be of interest to observe this event as it would tie down the parameters of the general model more narrowly. 
However, it will not give any further information on $q$. The 1922 outburst follows an impact on the outer disk, and these impact timings are not sensitive to $q$.

  Finally, let us comment on the effect of additional features introduced in the present study. First, the improvement in the timing of the outbursts with respect to Valtonen et al. (2010a) does improve the accuracy of the $q$ determination. Comparing set 4 with sets 1, 5 and 6 we see that the $\sigma$ of the distribution is greater by $\sim 10\%$ when the timing intervals are wider. If the angle between the disk axis and the primary spin axis varies as in Valtonen (2010a), the distribution is not centered on $q=1$ (set 10), unlike in our standard case. The solutions in Valtonen (2010a) were not numerous enough to detect this effect reliably.

 The effect of the $\chi_2$ on the $q$ distribution was calculated in three cases: first with the ``normal'' direction of the secondary spin $\chi_2$, then with the opposite spin, and finally with zero spin. The three distributions are different, showing that the spin-spin interaction has influenced the orbits (sets 1, 5 and 6), but the Gaussian parameters of the distributions are only marginally different from each other.

\section{The methods of observation of stars near to Sun and the possibilities to observe OJ287
in 2019}

The objects at small angular distances from the Sun (the estimated value for OJ287 is 8 to 12 degrees) are  difficult to observe due to high background caused by intense sunlight. The vignetting of the highly luminous solar disk enables a reduction of the background and  observion of stars near the Sun.
Recently two different methods have been used to observe stars at small angular distances from the Sun, namely the coronagraph method, and the helioscopic imager method.

\subsection{Coronagraph method}

The coronagraph has been in use already for a long time both in the ground based as well as in space based observations. The SOHO's LASCO C3 can serve as an example of a recent coronagraph in space (Morrill et al. 2006). The SOHO spacecraft has three coronagraphs (LASCO) onboard, two of which are still working (C2 and C3). 
We have used the publicly available images of these experiments to estimate the expected limiting magnitudes for this method. Estimating limiting magnitude for C2 is difficult because most of the LASCO C2 image is obscured by the solar corona and only a few stars are visible. LASCO C3 covers an area of 32 diameters of the Sun (i.e. about 16 degrees, hence the OJ287 position during the 2019 predicted flare would be covered) and  stars are clearly visible in the images, although a large part of the image is also obscured by the solar corona.

There are several problems which complicate the estimation of the limiting magnitude. The dominant one is the stray light which can be mistaken as background stars. There also several problems in determining the position and rotation of the spacecraft. One has to align the stars on the image from LASCO C3 and a star chart for the position (AAVSO charts can be used). For the elimination of the stay light, the video which is provided on the SOHO web pages was used. If there is a point source in five images in a row, we consider it a star. The rotation also complicates determination of the background stars. The best method here is to find a noticeably bright star and to align the image with the chart containing this star and its surroundings.

Pleiades are possibly the best objects to use, because the star cluster clearly defines the rotation. For our work we have used an image of Pleiades where  stars show up to the limiting magnitude of 10. It is compared with an the image from LASCO C3 (14.05.2010, Sun approaching Pleiades). The faintest stars of Pleiades detectable in the LASCO C3 image are of magnitude 8. Hence at best, we can obtain a limiting magnitude around 8 at the edges of the field where the corona is faint. Also the result strongly depends on the state of corona as during strong coronal mass ejections the limiting magnitude will be lower.

Probably a deeper magnitude could be achieved for a space based coronagraph with a larger aperture, but this would require an independent feasibility study as the previous coronagraphs were designed for solar studies, not for photometry of nearby stars.

In an independent study of the C3 limiting magnitude by Andrews (2000) using the same target (Pleiades cluster), deeper limits of magnitudes between 10 and 14 were achieved. As we would require a limiting magnitude of 15 or better for our timing measurement in July 2019, it is clear that a LASCO C3 type instrument will not be able to do the job.

\subsection{The imager method}

Another method recently used in a space experiment is the method of  heliospheric imager which is part of  the  SECCHI experiment (Howard et al. 2008, Eyles et al. 2009). The SECCHI experiment is onboard the STEREO space mission and consists of 
five telescopes, which together image the solar corona from the solar disk to beyond
 1 AU. These telescopes are: an extreme ultraviolet imager (EUVI: 1 - 1.7 solar radii), two traditional Lyot coronagraphs (COR1: 1.5 - 4 solar radii and COR2: 2.5 - 15 solar radii) and two new designs of heliospheric imagers (HI-1: 15 - 84 solar radii and HI-2: 66 - 318 solar radii). All the instruments use $2048\times2048$ pixel CCD arrays in a backside-in mode. The EUVI backside surface has been specially processed for EUV sensitivity, while the others have an anti-reflection coating applied.

The HI objectives, like the rest of the SECCHI suite, make visible light observations of CMEs and other structures as they transit from the corona and into the heliosphere. The HI package consists of two small, wide-angle telescope systems (HI-1 and HI-2) mounted on the side of each STEREO spacecraft, which together view the region between the Sun and the Earth. HI has no shutter mechanism, other than a one-shot door that protects the instrument from contamination during ground operations and the launch. Thus, an image is collected in a shutterless mode, in which the intensity at each pixel is an accumulation of the static scene and a smearing of the image during readout. This smearing can be removed on the ground.

The HI instrument concept was derived from the laboratory measurements of Buffington
et al. (1996) who determined the scattering rejection as a function of the number of occulters and the angle below the occulting edge. The concept is not unlike observing the night sky after the Sun has gone below the horizon.

While the specific concept used here has not been flown before, two other instruments
have flown which have validated the ability to measure the electron scattered component
against the strong zodiacal light and stellar background. The Zodiacal Light Photometer
(Pitz et al. 1976) on the Helios spacecraft, launched in 1974, and the Solar Mass
Ejection Imager (SMEI) instrument (Eyles et al. 2003), on the Coriolis spacecraft, launched in 2003 have demonstrated that a properly baffled instrument can detect CMEs (Tappin et al. 2003).

The HI-1 and HI-2 telescopes are directed to angles of about 13 degrees and 53 degrees from the principal axis of the instrument, which in turn is tilted upwards by 0.33 degrees to ensure that the Sun is sufficiently below the baffle horizon.

The novel heliospheric imagers achieve magnitude limits for stars of about 13 to 14 in a 40 minute exposure. The HI-1 imager covers the region of 7,5 to 24 degrees from Sun and is hence well suited to observe the OJ287 during the predicted 2019 event (OJ287 is expected to brighten from $\sim 14.3$ to $\sim 13$ in V-magnitude). A specific design for a dedicated space experiment to observe OJ287 optimizing the performace for stellar fotometry at angular distances around 10 degrees from the Sun may lead to a limiting magnitude increase to about 15.

Another possibility involves pointing 
Long Range Reconnaissance Imager (LORRI) in the `New Horizons' mission to Pluto
that consists of a telescope with a 20.8 centimeter 
aperture at OJ287 for one week in July 2019. In the $4\times4$ pixel binning mode the limiting magnitude in V is expected to be greater than 17 (Cheng et al. 2008).

\subsection{Secondary science}

The OJ287 heliospheric imager is expected to provide in addition to the OJ287 photometry  valuable data for other scientific fields such as monitoring of astrophysical targets near to Sun, and optical searches for optical counterparts to GRBs occurring at small angular distances from the Sun.

\section{Discussions and conclusions} 

We have shown that it is possible to test GR at the second PN order using the binary black hole system in OJ287. We find that GR can be confirmed with the one sigma accuracy of $30 \%$ using the currect observations and theoretical understanding of the system. One of the theoretical conditions is that the rotation axis of the accretion disk is at a constant angle with respect to the precessing spin axis of the primary black hole. It may be possible to verify this in future by studying the structure of the radio/X-ray jet in OJ287, and by theoretical studies of how the jet direction is determined when the two axes are not parallel to each other. With this proviso, we argue that it should be possible to 
test, in principle, the black hole no-hair theorem at the $10 \%$ level in the current decade by employing the binary black hole model of OJ287. 

We have also shown that the third order PN terms are too small to be detected in the OJ287 system. They depend on the exact value of the primary mass, at the $1 \%$ level, as well as on the spin of the secondary, for which it is difficult to find independent measurements at the required level of accuracy.  

We demonstrate that the testing at the above precision will require a certain amount of good luck 
in the sense that there should exist some yet unknown observations in the historical records at certain crucial time windows.
Also it is highly desirable to have space-based optical observations to monitor 
the impact outburst of OJ287 in 2019.
One possibility is to use the `New Horizons' mission to Pluto which by 2019 is already past Pluto. 

Employing a suitable 
solar observing mission to monitor OJ287 could be another option.
We have looked at the SOHO coronagraph images and find that the limiting magnitude there is 
$\sim 8$. We would need to get to magnitudes $\sim 15$ in blue or UV, and thus one would need to cover 
the innermost 5 degrees of the solar image instead of just the solar disk. A better case for the no-hair test would be a continuation or follow-up mission of STEREO which would for one week concentrate on OJ287 instead of solar flares.

Due to the fundamental nature of the test and the fact that the astrophysical systems associated 
with the other proposed tests are yet to be observed and likely to be plausible only in the next decade, it may not
be even be out of question to plan a small space mission 
to monitor the 2019 outburst and hence to test the black hole no-hair theorems.

\acknowledgments 
R.H. acknowledges the grants 102/09/0997 and 205/08/1207 by the GA CR and ME09027 by MSMT.

\end{document}